# Imaging Generalized Wigner Crystal States in a WSe$_2$/WS$_2$ Moiré Superlattice

*Hongyuan Li[1, 2, 3, 8], Shaowei Li[1, 3, 4, 8]\*, Emma C. Regan[1, 2, 3], Danqing Wang[1, 2], Wenyu Zhao[1], Salman Kahn[1, 3], Kentaro Yumigeta[5], Mark Blei[5], Takashi Taniguchi[6], Kenji Watanabe[7], Sefaattin Tongay[5], Alex Zettl[1, 3, 4], Michael F. Crommie[1, 3, 4]\* and Feng Wang[1, 3, 4]\*,*

[1]Department of Physics, University of California at Berkeley, Berkeley, CA, USA.

[2]Graduate Group in Applied Science and Technology, University of California at Berkeley, Berkeley, CA, USA.

[3]Materials Sciences Division, Lawrence Berkeley National Laboratory, Berkeley, CA, USA.

[4]Kavli Energy Nano Sciences Institute at the University of California Berkeley and the Lawrence Berkeley National Laboratory, Berkeley, CA, USA.

[5]School for Engineering of Matter, Transport and Energy, Arizona State University, Tempe, AZ, USA.

[6]International Center for Materials Nanoarchitectonics, National Institute for Materials Science, Tsukuba, Japan

[7]Research Center for Functional Materials, National Institute for Materials Science, Tsukuba, Japan

[8]These authors contributed equally: Hongyuan Li and Shaowei Li

**Abstract:**

The Wigner crystal state, first predicted by Eugene Wigner in 1934[1], has fascinated condensed matter physicists for nearly 90 years[2-14]. Studies of two-dimensional (2D) electron gases first revealed signatures of the Wigner crystal in electrical transport measurements at high magnetic fields[2-4]. More recently optical spectroscopy has provided evidence of generalized Wigner crystal states in transition metal dichalcogenide (TMDC) moiré superlattices[6-9]. Direct observation of the 2D Wigner crystal lattice in real space, however, has remained an outstanding challenge. Scanning tunneling microscopy (STM) in principle has sufficient spatial resolution to image a Wigner crystal, but conventional STM measurements can potentially alter fragile Wigner crystal states in the process of measurement. Here we demonstrate real-space imaging of 2D Wigner crystals in $WSe_2/WS_2$ moiré heterostructures using a novel non-invasive STM spectroscopy technique. We employ a graphene sensing layer in close proximity to the $WSe_2/WS_2$ moiré superlattice for Wigner crystal imaging, where local STM tunneling current into the graphene sensing layer is modulated by the underlying electron lattice of the Wigner crystal in the $WSe_2/WS_2$ heterostructure. Our measurement directly visualizes different lattice configurations associated with Wigner crystal states at fractional electron fillings of n = 1/3, 1/2, and 2/3, where n is the electron number per site. The n=1/3 and n=2/3 Wigner crystals are observed to exhibit a triangle and a honeycomb lattice, respectively, in order to minimize nearest-neighbor occupations. The n = 1/2 state, on the other hand, spontaneously breaks the original C3 symmetry and forms a stripe structure in real space. Our study lays a solid foundation toward the fundamental understanding of rich Wigner crystal states in $WSe_2/WS_2$ moiré heterostructures. Furthermore, this new STM technique is generally applicable to imaging novel correlated electron lattices in different van der Waals moiré heterostructures.

A Wigner crystal is the crystalline phase of electrons stabilized at low electron density where long-range Coulomb interactions dominate over quantum fluctuations in electron motion. The long pursuit of Wigner crystals[2-10] has motivated the study of 2D electron gases at high magnetic field where electron kinetic energy is quenched by degenerate Landau levels[15,16] and has led to the discovery of new quantum hall states[17,18]. Electrical transport signatures of Wigner crystal states have been reported in extremely clean GaAs/AlGaAs quantum wells[2,3] as well as graphene[4] at sufficiently low doping and high magnetic field. Signs of Wigner crystallization have also been detected for electrons trapped at the surface of liquid helium[11-14]. Recently, the discovery of moiré flat minibands in van der Waals heterostructures has opened a new route to realize Wigner crystal states at zero magnetic field. Several optical and conductance measurements have provided evidence of rich generalized Wigner crystal states in different TMDC moiré superlattices[6-9]. Direct observation of the real-space electron lattice in 2D, however, has remained challenging experimentally.

Real-space imaging of 2D Wigner crystals requires a measurement technique that satisfies several stringent requirements. It must (1) have sufficient spatial resolution to resolve the electron lattice, (2) have sufficient sensitivity to detect the presence of single electrons in the lattice, (3) be adequately non-invasive to not destroy fragile Wigner crystal states. The last two requirements conflict with each other since strong coupling to the Wigner crystal is required for high sensitivity, whereas weak coupling is required to avoid strongly perturbing fragile states. For example, conventional STM measurements have excellent spatial resolution and charge sensitivity but can be highly invasive since inevitable tip-gating effects at finite tip bias can destroy the delicate electron lattice of the Wigner crystal. In this work, we utilize a novel STM measurement scheme that strikes a balance between these two contradictory requirements, thus enabling real-space imaging of the n=2/3, n=1/2, and n=1/3 2D Wigner crystal states in WSe$_2$/WS$_2$ moiré heterostructures.

Our new STM scheme employs a specially designed van der Waals heterostructure as illustrated in Fig. 1a (see Methods for the sample fabrication details). It integrates a gated WSe$_2$/WS$_2$ moiré heterostructure and a top graphene monolayer sensing layer that are separated by a hexagonal boron nitride (hBN) layer with a thickness $d_t$ = 5nm, chosen to be smaller than the moiré lattice constant ($L_M$ = 8nm). This separation is small enough that the STM tip and graphene sensing layer can efficiently couple to individual moiré electrons in the WSe$_2$/WS$_2$ superlattice, but it is large enough that the tip and graphene layer remain non-invasive with respect to the delicate Wigner crystal states. STM tunneling current into the graphene sensing layer can be modulated by the charge states of different moiré sites in the WSe$_2$/WS$_2$ superlattice through local Coulomb blockade effects[19]. This technique allows us to detect the local charge distribution in the WSe$_2$/WS$_2$ heterostructure and to image the embedded Wigner crystal lattice.

Fig. 1b shows a typical large-scale topography image measured on the top graphene surface. The top graphene and hBN layers cover the WSe$_2$/WS$_2$ heterostructure conformally and inherit the topography of the 3D reconstructed moiré superlattice[20] below. Fig. 1c shows a zoom-in topographic image corresponding to the red dashed box area in Fig. 1b. A red rhombus labels the primitive cell of the moiré superlattice with the four high points corresponding to AA stacking regions and the two inequivalent low points corresponding to distinct AB stacking regions (denoted AB$_1$ and AB$_2$)[20]. The measured moiré lattice constant is $L_M$ = 8nm, yielding the twist angle between the WSe$_2$ and WS$_2$ layers $\theta \sim 0°$ through the formula $L_M = \frac{a}{\sqrt{\delta^2 + \theta^2}}$, where $\delta = (a - a')/a$ is the lattice mismatch, a = 3.153 Å and a' = 3.28 Å are the atomic lattice constants of the WS$_2$ and WSe$_2$, respectively[21].

We implement dual gates in our van der Waals heterostructure devices (Fig. 1a) with the top monolayer graphene acting as both sensing layer and top gate and the silicon substrate acting as back gate. The top gate dielectric is defined by the top hBN flake ($d_t$ = 5nm) while the bottom gate dielectric is defined by a combination of SiO$_2$ ($d_{SiO2}$ = 285nm) and hBN ($d_b$ = 70nm). The carrier densities in the TMDC moiré heterostructure and the top graphene can be controlled independently via the top gate voltage, $V_{TG}$, and bottom gate voltage, $V_{BG}$. In this study we mainly focus on the electron-doped regime of the WSe$_2$/WS$_2$ heterostructure.

For $V_{TG}$ = 0, the Fermi level is within the band gap for the WSe$_2$/WS$_2$ heterostructure (see illustration in Fig. 1d). In this case, tuning $V_{BG}$ dopes charge carriers exclusively into the graphene layer. Fig. 1e shows a 2D plot of the STM differential conductivity (dI/dV) spectra of graphene at $V_{TG}$ = 0 for different values of $V_{BG}$. (See SI for individual dI/dV spectra at different gate voltages). The dispersive feature labeled by the white dashed line shows the evolution of the graphene charge neutral point (CNP) in response to the electrostatic doping from $V_{BG}$[22-24]. The persistent gap near $V_{bias}$ = 0 arises from an inelastic tunneling gap that occurs at all gate voltages. This inelastic tunneling gap causes the graphene CNP curve to abruptly shift as it shifts over the zero-bias region[22,24].

We are able to dope electrons into the WSe$_2$/WS$_2$ heterostructure by applying a positive $V_{TG}$ such that the Fermi level of the WSe$_2$/WS$_2$ heterostructure lies near the conduction band edge (see illustration in Fig. 1f). Here we choose $V_{TG}$ ~ 0.5V so that the WSe$_2$/WS$_2$ heterostructure can be electron doped while the graphene sensing layer remains close to charge neutral. The reason for doing this is that the resulting small density of states for graphene provides the highest sensitivity for imaging Wigner crystal states in the moiré superlattice. Charge neutral graphene also has less of a screening effect on the moiré electron-electron interactions due to the long screening length of Dirac electrons at the CNP[25]. Fig. 1g shows the resulting dI/dV tunneling spectra into the graphene sensing layer as a function of $V_{BG}$ at a fixed $V_{TG}$ = 0.53V. This panel corresponds to the same {$V_{BG}$, $V_{bias}$} phase space outlined by the dashed white box in Fig.1e, but for nonzero $V_{TG}$.

Fig. 1g shows that the graphene is hole-doped at $V_{BG}$ < 7V, while it is impossible for the WSe$_2$/WS$_2$ heterostructure to be hole-doped under these conditions given the band alignment shown in Fig. 1f. The graphene hole doping leads to dispersive movement of the graphene CNP at $V_{BG}$ < 7V (denoted by the white dashed line in Fig. 1g). Electron doping for $V_{BG}$ > 7V, however, leads to very different behavior. In a non-interacting single-particle picture, the electron doping would occur predominantly in the WSe$_2$/WS$_2$ heterostructure because its density of states (DOS) is orders of magnitude larger than the graphene DOS at the Dirac point. Therefore, one would expect the graphene Fermi energy to stay fixed near the Dirac point, as illustrated by the vertical dashed line at $V_{BG}$ > 7V. Experimentally, however, we observe a non-trivial shift of the graphene CNP with respect to the Fermi energy at different $V_{BG}$ values. The graphene layer undergoes electron doping when the WSe$_2$/WS$_2$ heterostructure experiences fractional filling of the moiré superlattice with n=1/3, 1/2, 2/3, and 1 (black dashed lines in Fig. 1g). Figure 1h shows a vertical line-cut of the gate-dependent dI/dV spectra at $V_{bias}$ = 0.1V, which shows clear peaks at these fractional fillings. These features signify the correlated gaps in the WSe$_2$/WS$_2$ Mott insulator state at n=1 as well as the generalized Wigner crystal insulator states at n=1/3, 1/2, 2/3. This is due to the fact that the correlated gaps make the WSe$_2$/WS$_2$ heterostructure electronically incompressible and so electrons are electrostatically forced into the graphene sensing layer. Similar effects have been observed for capacitance and single-electron-transistor measurements of electronic compressibility in different van der Waals heterostructure systems[26-30]. Our STM configuration thus provides a new technique for mapping the local electronic compressibility of correlated insulating states in moiré superlattices.

Real-space imaging of the 2D electron lattice of the Mott insulator and Wigner crystal states was performed through 2D dI/dV mapping of the graphene sensing layer, as illustrated in the measurement scheme of Fig. 2a. The Mott and Wigner crystal states form periodic electron lattices in the $WSe_2/WS_2$ moiré heterostructure that couple to the graphene sensing layer and STM tip through long-range Coulomb interactions. The tunnel current between the STM tip and the graphene layer will vary spatially depending on the charge state of the $WSe_2/WS_2$ moiré site below the STM tip. As a result, the electron lattices of the Mott and Wigner crystal states can be imaged as periodic lattice structures in 2D dI/dV mappings of the graphene sensing layer.

Fig. 2b displays the topographic image of a typical region of the $WSe_2/WS_2$ moiré superlattice as seen by scanning the top graphene layer. The triangular lattice formed by the $AB_1$ stacking sites have been marked with solid red dots. This region exhibits a lattice that is free of distortion or atomic defects, an essential condition for observing Mott and Wigner crystal states with long-range order.

Fig. 2c shows a dI/dV mapping of the graphene layer when the $WSe_2/WS_2$ moiré heterostructure is in the n = 1 Mott insulator state ($V_{bias}$ = 160mV, $V_{BG}$ = 30V, and $V_{TG}$ = 0.53 V; see Methods for more measurement details). A highly ordered triangular lattice of bright features can be clearly observed that corresponds to the $AB_1$ stacking sites of the moiré superlattice. Such $AB_1$ stacking sites are illustrated as red dots in Fig. 2b. Since the $AB_1$ and $AB_2$ stackings sites are similar in topography, the centering of the bright features in Fig. 1c on the $AB_1$ stacking sites implies that these features do not originate from topography, but rather from the underlying electron lattice of the Mott insulator state. This is consistent with previous work showing that conduction flat band electrons in the $WSe_2/WS_2$ moiré heterostructure are localized at one of the AB stacking sites[19]. Fig. 2d displays the fast Fourier transform (FFT) image of the dI/dV map in Fig. 2c, showing sharp diffraction points associated with the electron lattice of the Mott insulator state. The reciprocal unit vectors of the moiré superlattice are marked by green dots in the FFT image and are seen to overlap perfectly with the lowest order diffraction points of the Mott insulator electron lattice.

We next imaged the generalized Wigner crystal states at fractional fillings. Fig. 2e shows the dI/dV mapping of the n = 2/3 generalized Wigner crystal state ($V_{bias}$ = 160mV, $V_{BG}$ = 21.8V, $V_{TG}$ = 0.458 V). FFT filtering has been performed on this and subsequent Wigner crystal images in Fig. 2 to suppress periodic features associated with the moiré superlattice (i.e., the green dots in Fig. 2f. See SI for unfiltered images). The n = 2/3 dI/dV map exhibits a honeycomb lattice with lattice constant $\sqrt{3}L_M$. This is consistent with 2/3 of the available $AB_1$ sites being filled with electrons (i.e., the solid red dots located in bright regions of the image) and the other 1/3 $AB_1$ sites being empty (i.e., the open red circles located in dark regions of the image) so as to minimize total nearest-neighbor interactions. Such a honeycomb lattice matches previous predictions[6-9] and confirms the existence of generalized Wigner crystal states where moiré electrons are stabilized by long-range Coulomb interactions and exhibit well-defined 2D crystalline order. The corresponding FFT image (Fig. 2f) demonstrates the emergence of a new lattice: six sharp diffraction points associated with the generalized Wigner crystal lattice appear *inside* the reciprocal unit vectors (green dots) of the moiré superlattice.

Figs. 2g and 2h show the dI/dV map and corresponding FFT image, respectively, of the n=1/3 generalized Wigner crystal state ($V_{bias}$ = 130mV, $V_{BG}$ = 14.9V, $V_{TG}$ = 0.458 V). The real-space image demonstrates a new triangular electron lattice associated with a Wigner crystal state where 1/3 of the available $AB_1$ sites are filled with electrons (solid red dots) and the other 2/3 $AB_1$ sites are empty (open red circles). The FFT image shows a clear diffraction pattern of the generalized Wigner crystal state with a lattice constant of $\sqrt{3}L_M$. We note that the diffraction pattern of the n = 2/3 and n = 1/3 states are nearly identical because both states share the same primitive cell and are linked by a particle-hole transformation.

Fig. 2i shows a dI/dV map of the n = 1/2 generalized Wigner crystal state ($V_{bias}$ = 125mV, $V_{BG}$ = 18.7V, and $V_{TG}$ = 0.458 V). The image reveals unambiguously that the C3 symmetry of the host moiré superlattice is spontaneously broken for this generalized Wigner crystal state. The n = 1/2 state features a stripe symmetry with electrons (solid red dots) filling the $AB_1$ sites in alternating lines (empty sites are again marked by open red circles). The lattice constants of this stripe phase are $L_M$ and $\sqrt{3}L_M$ along the parallel and perpendicular directions, respectively. The corresponding FFT image in Fig. 2j shows a rhombus-like reciprocal unit vectors, further confirming the broken symmetry of this stripe phase for the n = 1/2 generalized Wigner crystal state.

The n=1/2 generalized Wigner crystal state is predicted to be highly degenerate, with multiple electron lattice configurations having the same energy in the case of only nearest-neighbor interactions[7]. The spontaneous broken symmetry of the n = 1/2 state might therefore be governed by higher-order effects, such as next-nearest-neighbor interactions and/or accidental strain in the lattice. Experimentally we here found that the n = 1/2 state is more fragile than the n=1/3 and n=2/3 states. A well-defined generalized Wigner crystal stripe phase is present only in a very narrow parameter space of $V_{bias}$ and $V_{BG}$. The n = 1/2 state is also more sensitive to local inhomogeneity, as reflected by the disordering of the stripe electron lattice near the right edge of the image in Fig. 2i as compared to the more defect-free n=2/3 (Fig. 2c) and n=1/3 (Fig. 2e) states. Further studies of the generalized Wigner crystal electron lattice at n = 1/2 could potentially lead to a better understanding of the competition of different quantum phases controlled by long-range Coulomb interactions.

We last discuss the imaging mechanism underlying the dI/dV mapping of generalized Wigner crystal electron lattices in the presence of a graphene sensing layer. As illustrated in Fig. 2a, the STM tunneling current into the graphene layer can be coupled to moiré electrons below the tip through long-range Coulomb interactions. This coupling can affect the tunnel current in two different ways: (1) the localized moiré electrons can induce local band bending in the graphene sensing layer, thus changing the graphene local density of states and hence the dI/dV signal. (2) application of $V_{bias}$ can discharge the moiré electron right below the tip once $V_{bias}$ exceeds a threshold value. This mechanism is helped by the fact that electrical screening by the top monolayer graphene is weak when its Fermi level is close to the Dirac point. A resulting moiré electron discharging event could then lead to a sudden increase in the STM tunneling current due to the elimination of the Coulomb blockade effect, hence contributing to a stronger dI/dV signal[19].

To distinguish between these two mechanisms we have systematically examined how the dI/dV maps evolve as $V_{bias}$ is changed. Figs. 3a-e show dI/dV maps of the n = 2/3 Wigner crystal state as $V_{bias}$ is increased from 130mV to 190mV in 15mV steps. No FFT filtering was performed on these images. The honeycomb electron lattice associated with the n = 2/3 generalized Wigner crystal state is not so clearly seen in the map at $V_{bias}$ = 130 mV (Fig. 3a), but emerges when $V_{bias}$ is increased to 145 mV (Fig. 3b). The dominant features are the bright dots centered on the $AB_1$ stacking sites. These features expand with increased $V_{bias}$ (Fig. 3c) and ultimately form ring-like features (Figs. 3d,e). Such behavior (i.e. expanding rings with increased tip bias) is characteristic of tip-induced electrical discharging rings[19,31-34] and occurs because electrical discharging for larger tip-electron distances requires larger tip biases. This indicates that mechanism (2) discussed above is the dominant contrast mechanism for imaging Wigner crystal states in our dI/dV maps. The STM tip locally discharges the moiré electron localized at the $AB_1$ site closest to the tip apex once $V_{bias}$ is large enough and the tip-electron distance is short enough. This enables discharge features centered around *filled* $AB_1$ sites to be observed in dI/dV maps of the graphene sensing layer.

In conclusion, we have developed a new STM imaging technique that combines high spatial resolution and sensitivity with minimal perturbation to the probed electronic system. This technique has enabled us to directly image the delicate 2D electron lattices of generalized Wigner crystals in real space. Our technique should be generally applicable to a wide variety of van der Waals moiré heterostructures and provides a powerful new tool for imaging real-space electron configurations of novel correlated quantum phases in 2D systems.

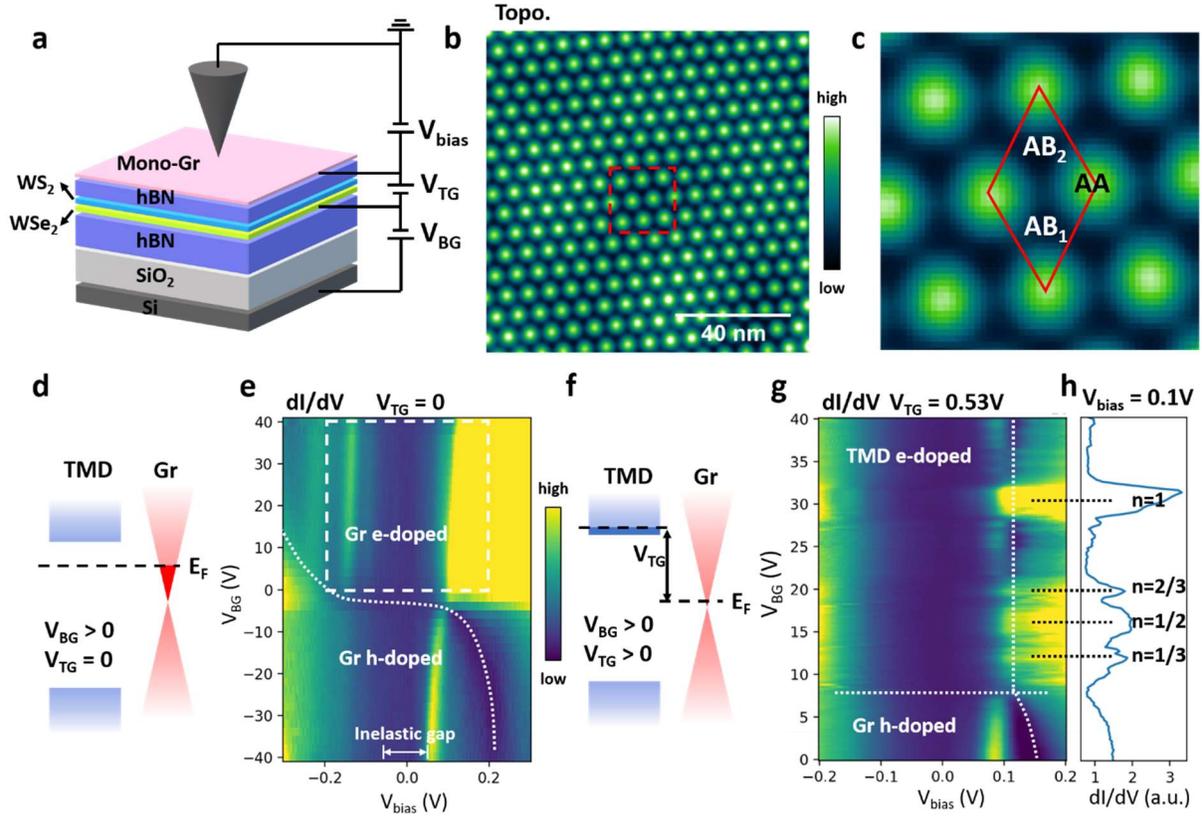

**Figure 1. STM measurement of Wigner crystal states in a dual-gated WSe$_2$/WS$_2$ moiré superlattice. a.** Schematic of the dual-gated WSe$_2$/WS$_2$ moiré heterostructure device. The top hBN thickness (5nm) is slightly smaller than the moiré lattice constant (8nm). Top gate ($V_{TG}$) and bottom gate ($V_{BG}$) voltages are applied to separately control the carrier density in the WSe$_2$/WS$_2$ heterostructure as well as the top graphene sensing layer. **b.** A typical large-scale topography image measured on the top graphene surface. $V_{bias}$ = 180mV and I = 300 pA. **c.** Zoom-in image of the red dashed box in (**b**). The red rhombus labels a primitive cell. Peaks correspond to AA stacking regions and the two inequivalent low points correspond to distinct AB stacking regions (denoted AB$_1$ and AB$_2$). **d.** Schematic of the heterostructure band alignment and Fermi levels for $V_{TG}$ = 0 and $V_{BG}$ > 0. At zero $V_{TG}$, the Fermi level of the WSe$_2$/WS$_2$ heterostructure is located in the band gap. **e.** $V_{BG}$-dependent dI/dV spectra measured on the graphene sensing layer over an AA stacking site for $V_{TG}$ = 0. The dispersive feature marked by the white dotted curve shows the evolution of the graphene charge neutral point (CNP) induced by electrostatic doping from $V_{BG}$. The persistent gap near $V_{bias}$ = 0 arises from an inelastic tunneling gap that exists at all gate voltages. Due to this inelastic tunneling gap, the graphene CNP curve shows an abrupt shift as it shifts over the zero-bias region. The tip height was set by the following parameters: $V_{bias}$ = -300mV and I = 100 pA. **f.** Schematic of the heterostructure band alignment and Fermi levels for $V_{TG}$ > 0 and $V_{BG}$ > 0. Application of an appropriate positive $V_{TG}$ allows the Fermi level of the WSe$_2$/WS$_2$ heterostructure to be lifted into the conduction band. **g.** $V_{BG}$-dependent dI/dV spectra measured on the graphene sensing layer over an AA stacking site for $V_{TG}$ = 0.53V. This is a zoom-in of the electron doped regime corresponding to the phase space denoted by the white dashed box in **e**. The graphene is hole-doped in the region below the horizontal dashed line ($V_{BG}$ < 7V), and the WSe$_2$/WS$_2$ is electron-doped in the region above it ($V_{BG}$ > 7V). The vertical white dash curve indicates the expected movement of the graphene CNP for a non-interacting picture. Significant electron doping of the graphene layer takes place at n = 1/3, 1/2, 2/3, and 1 (denoted by horizontal black dashed lines in (**g**)). The tip height was set by the following parameters: $V_{bias}$ = -200mV and I = 100 pA. **h**. Vertical line-cut through the $V_{BG}$-dependent dI/dV spectra in (**g**) at $V_{bias}$ = 0.1V shows peaks at n=1, 2/3, 1/2, and 1/3.

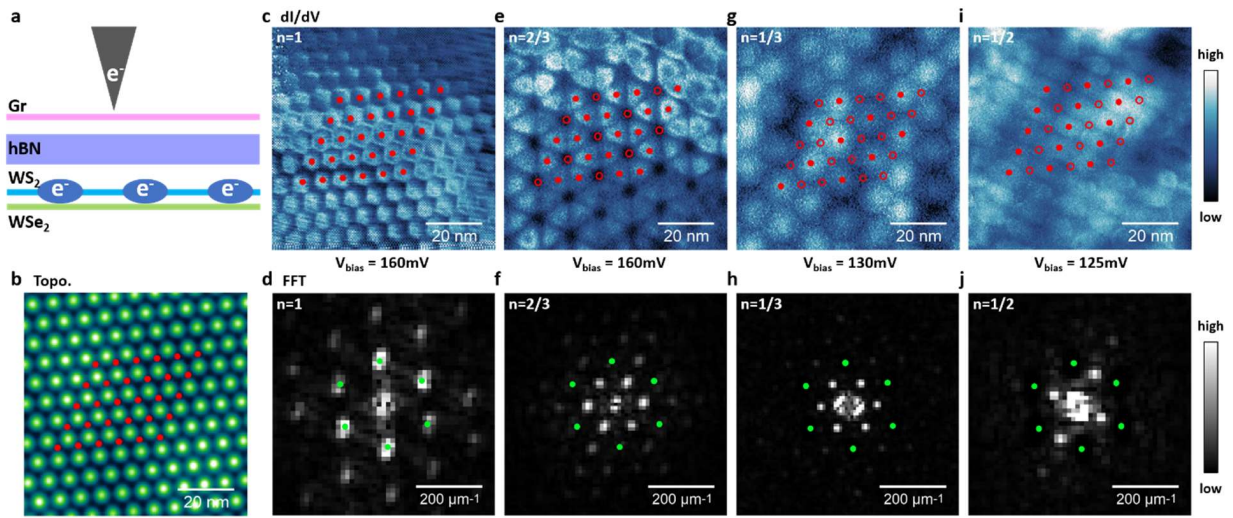

**Figure 2. Imaging Mott and generalized Wigner crystal states. a**. Schematic shows imaging of correlated states in a WSe$_2$/WS$_2$ moiré superlattice beneath a graphene sensing layer. dI/dV maps are acquired at the top graphene surface. **b**. A typical STM topographic image of the moiré superlattice shows a perfect lattice without distortion or defects. The triangular lattice formed by the AB$_1$ stacking sites is marked by solid red dots. **c**. dI/dV map of the n = 1 Mott insulator ($V_{bias}$ = 160mV, $V_{BG}$ = 30V, $V_{TG}$ = 0.53 V). The triangular lattice formed by the AB$_1$ stacking sites is labeled with red dots. **d**. Fast Fourier transform (FFT) of the image shown in (**c**). The reciprocal unit vectors of the moiré superlattice are labeled by green dots. **e-j**. dI/dV maps of the generalized Wigner crystal states for different electron fillings and their corresponding FFT images: **e**. dI/dV map of n = 2/3 state ($V_{bias}$ = 160mV, $V_{BG}$ = 21.8V, $V_{TG}$ = 0.458 V). **f**. FFT of n = 2/3 state shown in (**e**). **g**. dI/dV map of n = 1/3 state ($V_{bias}$ = 130mV, $V_{BG}$ = 14.9V, $V_{TG}$ = 0.458 V). **h**. FFT of n = 1/3 state shown in (**g**). **i**. dI/dV map of n = 1/2 state ($V_{bias}$ = 125mV, $V_{BG}$ = 18.7V, $V_{TG}$ = 0.458 V). **j**. FFT of n = 1/2 state shown in (**i**). FFT filtering was performed in the Wigner crystal images (**e-j**) to suppress the periodic features associated with the moiré superlattice (i.e., green dots in (**f,h,j**); see SI for unfiltered images). Electron-filled AB$_1$ sites are labeled with solid red dots and the empty AB$_1$ sites are labeled with open red circles in the real-space images (**e,g,i**). The locations of the reciprocal unit vectors of the moiré superlattice are marked by green dots in (**f,h,j**).

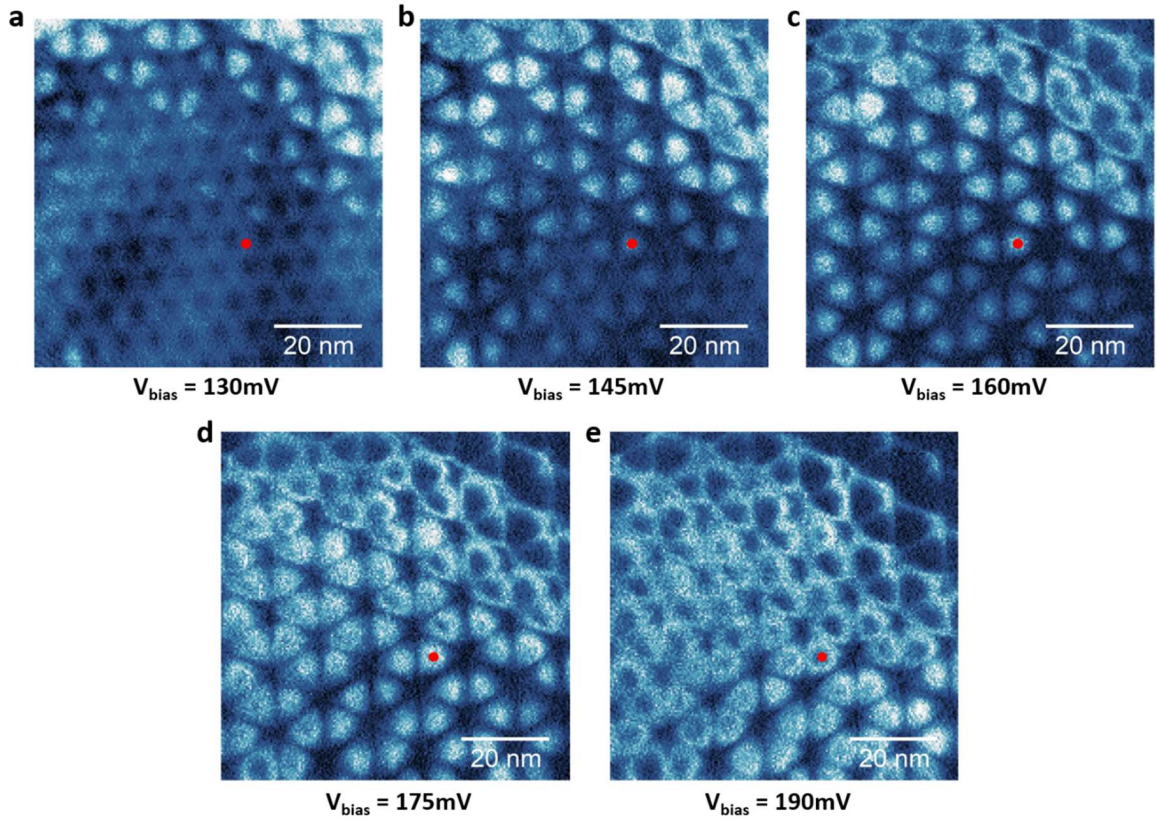

**Figure 3. Evolution of dI/dV maps for the n = 2/3 state with increased $V_{bias}$.** dI/dV maps of the n = 2/3 generalized Wigner crystal state measured for (**a**) $V_{bias}$ = 130mV, (**b**) $V_{bias}$ = 145mV, (**c**) $V_{bias}$ = 160mV, (**d**) $V_{bias}$ = 175mV, and (**e**) $V_{bias}$ = 190mV. Gate voltage parameters: $V_{BG}$ = 21.8V, $V_{TG}$ = 0.458 V. All five maps are measured in the same region and no filtering has been performed. The map in (**c**) ($V_{bias}$ = 160mV) shows the same data as Fig. 2e, but with no filtering. The red dot labels one typical electron-filled $AB_1$ site where a discharging ring can be observed that gets larger with increased $V_{bias}$ (a common characteristic of discharging phenomena).

**Corresponding Author**

* Email: swli@berkeley.edu (S.L.), crommie@physics.berkeley.edu (M.C.) and fengwang76@berkeley.edu (F.W.).

**Corresponding Author**

* Email: swli@berkeley.edu (S.L.), crommie@physics.berkeley.edu (M.C.) and fengwang76@berkeley.edu (F.W.).

**Author Contributions**

M.C. and F.W. conceived the project. H.L., S.L. performed the STM measurement, H.L., E.R., D.W., W.Z., and S.K. fabricated the heterostructure device and performed the SHG measurement. K.Y., M.B. and S.T. grew $WSe_2$ and $WS_2$ crystals. K.W. and T.T. grew the hBN single crystal. All authors discussed the results and wrote the manuscript.

**Notes**

The authors declare no financial competing interests.

**ACKNOWLEDGMENT**

This work was primarily funded by the U.S. Department of Energy, Office of Science, Office of Basic Energy Sciences, Materials Sciences and Engineering Division under Contract No. DE-AC02-05-CH11231 (van der Waals heterostructure program KCFW16) (device electrode preparation and STM spectroscopy). Support was also provided by the US Army Research Office under MURI award W911NF-17-1-0312 (device layer transfer), and by the National Science Foundation Award DMR-1807233 (surface preparation). S.T acknowledges support from DOE-SC0020653, NSF DMR 2111812, NSF DMR 1552220, NSF 2052527, DMR 1904716, and NSF CMMI 1933214 for $WSe_2$ and $WS_2$ bulk crystal growth and analysis. K.W. and T.T. acknowledge support from the Elemental Strategy Initiative conducted by the MEXT, Japan, Grant Number JPMXP0112101001, JSPS KAKENHI Grant Number JP20H00354 and the CREST(JPMJCR15F3), JST for bulk hBN crystal growth and analysis. E.C.R. acknowledges support from the Department of Defense (DoD) through the National Defense Science & Engineering Graduate Fellowship (NDSEG) Program. S.L. acknowledges support from Kavli ENSI Heising Simons Junior Fellowship. The authors also thank M. H. Naik for sharing unpublished theoretical simulation data on WSe$_2$/WS$_2$ moiré superlattice.

**Methods**

**Sample fabrication:** The encapsulated WSe$_2$/WS$_2$ moiré heterostructure stack was fabricated using the micro-mechanical stacking technique[35]. A poly(propylene) carbonate (PPC) film stamp was used to pick up all exfoliated 2D material flakes. The 2D material layers in the main heterostructure region were picked up in the following order: bottom hBN, WSe$_2$, WS$_2$, top hBN, and then monolayer graphene. A graphite layer was also picked up between the WS$_2$ and the top hBN to serve as a contact electrode of the WSe$_2$/WS$_2$ heterostructure. The PPC film together with the stacked sample was then peeled, flipped over, and transferred onto a Si/SiO$_2$ substrate (SiO$_2$ thickness 285nm). The PPC layer was subsequently removed using ultrahigh vacuum annealing at 230 °C, resulting in an atomically-clean heterostructure suitable for STM measurements. A 50nm Au and 5nm Cr metal layer was then evaporated through a shadow mask to form the electric contact.

**STM and STS measurement:** STM and STS measurements were performed using a Pt/Ir etched tip at T = 5.4 K and pressure = $2 \times 10^{-10}$ Torr. The STS (dI/dV spectra and mapping) were performed with a tip bias modulation of 15mV amplitude and 500~900 Hz frequency. The dI/dV mapping of the Mott-insulator state in Fig. 2c was performed under open-loop conditions with the tip height set by the following parameters: V$_{bias}$ = 180mV and I = 300 pA. dI/dV mappings of the generalized Wigner crystals states were performed under constant-current mode if not further specified.

**Supplementary Materials**

**Data availability**

The data supporting the findings of this study are included in the main text and in the Supplementary Information files, and are also available from the corresponding authors upon reasonable request.

**Reference**

1   Wigner, E. On the interaction of electrons in metals. *Physical Review* **46**, 1002 (1934).
2   Goldman, V., Santos, M., Shayegan, M. & Cunningham, J. Evidence for two-dimentional quantum Wigner crystal. *Physical review letters* **65**, 2189 (1990).
3   Jang, J., Hunt, B. M., Pfeiffer, L. N., West, K. W. & Ashoori, R. C. Sharp tunnelling resonance from the vibrations of an electronic Wigner crystal. *Nature Physics* **13**, 340-344 (2017).
4   Zhou, H., Polshyn, H., Taniguchi, T., Watanabe, K. & Young, A. Solids of quantum Hall skyrmions in graphene. *Nature Physics* **16**, 154-158 (2020).
5   Shapir, I. *et al.* Imaging the electronic Wigner crystal in one dimension. *Science* **364**, 870-875 (2019).
6   Regan, E. C. *et al.* Mott and generalized Wigner crystal states in WSe 2/WS 2 moiré superlattices. *Nature* **579**, 359-363 (2020).
7   Jin, C. *et al.* Stripe phases in WSe2/WS2 moir\'e superlattices. *arXiv preprint arXiv:2007.12068* (2020).
8   Xu, Y. *et al.* Correlated insulating states at fractional fillings of moiré superlattices. *Nature* **587**, 214-218 (2020).
9   Huang, X. *et al.* Correlated insulating states at fractional fillings of the WS 2/WSe 2 moiré lattice. *Nature Physics*, 1-5 (2021).
10  Deshpande, V. V. & Bockrath, M. The one-dimensional Wigner crystal in carbon nanotubes. *Nature Physics* **4**, 314-318 (2008).
11  Crandall, R. & Williams, R. Crystallization of electrons on the surface of liquid helium. *Physics Letters A* **34**, 404-405 (1971).
12  Williams, R., Crandall, R. & Willis, A. Surface states of electrons on liquid helium. *Physical Review Letters* **26**, 7 (1971).
13  Grimes, C. & Adams, G. Evidence for a liquid-to-crystal phase transition in a classical, two-dimensional sheet of electrons. *Physical Review Letters* **42**, 795 (1979).

14 Williams, F. Collective aspects of charged-particle systems at helium interfaces. *Surface Science* **113**, 371-388 (1982).
15 Lam, P. K. & Girvin, S. Liquid-solid transition and the fractional quantum-Hall effect. *Physical Review B* **30**, 473 (1984).
16 Levesque, D., Weis, J. & MacDonald, A. Crystallization of the incompressible quantum-fluid state of a two-dimensional electron gas in a strong magnetic field. *Physical Review B* **30**, 1056 (1984).
17 Tsui, D. C., Stormer, H. L. & Gossard, A. C. Two-dimensional magnetotransport in the extreme quantum limit. *Physical Review Letters* **48**, 1559 (1982).
18 Klitzing, K. v., Dorda, G. & Pepper, M. New method for high-accuracy determination of the fine-structure constant based on quantized Hall resistance. *Physical review letters* **45**, 494 (1980).
19 Li, H. *et al.* Imaging local discharge cascades for correlated electrons in WS2/WSe2 moir\'e superlattices. *arXiv preprint arXiv:2102.09986* (2021).
20 Li, H. *et al.* Imaging moiré flat bands in three-dimensional reconstructed WSe 2/WS 2 superlattices. *Nature Materials*, 1-6 (2021).
21 Schutte, W., De Boer, J. & Jellinek, F. Crystal structures of tungsten disulfide and diselenide. *Journal of Solid State Chemistry* **70**, 207-209 (1987).
22 Zhang, Y. *et al.* Giant phonon-induced conductance in scanning tunnelling spectroscopy of gate-tunable graphene. *Nature Physics* **4**, 627-630 (2008).
23 Jung, S. *et al.* Evolution of microscopic localization in graphene in a magnetic field from scattering resonances to quantum dots. *Nature Physics* **7**, 245-251 (2011).
24 Decker, R. *et al.* Local electronic properties of graphene on a BN substrate via scanning tunneling microscopy. *Nano letters* **11**, 2291-2295 (2011).
25 Wong, D. *et al.* Spatially resolving density-dependent screening around a single charged atom in graphene. *Physical Review B* **95**, 205419 (2017).
26 Yang, F. *et al.* Experimental determination of the energy per particle in partially filled Landau levels. *arXiv preprint arXiv:2008.05466* (2020).
27 Li, T. *et al.* Charge-order-enhanced capacitance in semiconductor moir\'e superlattices. *arXiv preprint arXiv:2102.10823* (2021).
28 Tomarken, S. L. *et al.* Electronic compressibility of magic-angle graphene superlattices. *Physical review letters* **123**, 046601 (2019).
29 Zondiner, U. *et al.* Cascade of phase transitions and Dirac revivals in magic-angle graphene. *Nature* **582**, 203-208 (2020).
30 Pierce, A. T. *et al.* Unconventional sequence of correlated Chern insulators in magic-angle twisted bilayer graphene. *arXiv preprint arXiv:2101.04123* (2021).
31 Pradhan, N. A., Liu, N., Silien, C. & Ho, W. Atomic scale conductance induced by single impurity charging. *Physical review letters* **94**, 076801 (2005).
32 Brar, V. W. *et al.* Gate-controlled ionization and screening of cobalt adatoms on a graphene surface. *Nature Physics* **7**, 43-47 (2011).
33 Wong, D. *et al.* Characterization and manipulation of individual defects in insulating hexagonal boron nitride using scanning tunnelling microscopy. *Nature nanotechnology* **10**, 949-953 (2015).
34 Teichmann, K. *et al.* Controlled charge switching on a single donor with a scanning tunneling microscope. *Physical review letters* **101**, 076103 (2008).
35 Wang, L. *et al.* One-dimensional electrical contact to a two-dimensional material. *Science* **342**, 614-617 (2013).

# Supplementary Information for
# Imaging Generalized Wigner Crystal States in a WSe$_2$/WS$_2$ Moiré Superlattice

*Hongyuan Li[1,2,3,8], Shaowei Li[1,3,4,8]\*, Emma Regan[1,2,3], Danqing Wang[1,2], Wenyu Zhao[1], Salman Kahn[1,3], Kentaro Yumigeta[5], Mark Blei[5], Takashi Taniguchi[6], Kenji Watanabe[7], Sefaattin Tongay[5], Alex Zettl[1,3,4], Michael F. Crommie[1,3,4]\* and Feng Wang[1,3,4]\*,*

[1]Department of Physics, University of California at Berkeley, Berkeley, CA, USA.

[2]Graduate Group in Applied Science and Technology, University of California at Berkeley, Berkeley, CA, USA.

[3]Materials Sciences Division, Lawrence Berkeley National Laboratory, Berkeley, CA, USA.

[4]Kavli Energy Nano Sciences Institute at the University of California Berkeley and the Lawrence Berkeley National Laboratory, Berkeley, CA, USA.

[5]School for Engineering of Matter, Transport and Energy, Arizona State University, Tempe, AZ, USA.

[6]International Center for Materials Nanoarchitectonics, National Institute for Materials Science, Tsukuba, Japan

[7]Research Center for Functional Materials, National Institute for Materials Science, Tsukuba, Japan

[8]These authors contributed equally: Hongyuan Li and Shaowei Li

1. **Individual dI/dV spectra**

Fig. S1a displays the evolution of the dI/dV spectra for $V_{TG} = 0$ and different values of $V_{BG}$. We observe that the Dirac point (characterized by a dip in the LDOS and marked by grey arrows) shifts from positive bias to negative bias voltage with increased $V_{BG}$, corresponding to a change from the graphene hole-doped regime to the graphene electron-doped regime. The small tunneling current close to zero bias voltage is due to reduced tunneling from graphene electrons at the K and K' points of the graphene band structure[1] (phonon-induced inelastic tunneling additionally causes rises in tunnel current that can result in gap-like features in graphene dI/dV spectra[1]).

Fig. S1b displays the dI/dV spectra at $V_{TG} = 0.53V$ for different $V_{BG}$. The $WS_2/WSe_2$ moiré heterostructure is conducting at $V_{BG} = 6V$, $25V$, and $37V$. The dI/dV spectra for these backgate voltages remain almost constant and correspond to charge neutral or a weak hole doping. These spectra are comparable to the spectrum at $V_{BG} = -6V$ for $V_{TG}=0V$ (shown in Fig. S1a). The dI/dV spectra change significantly at $V_{BG} = 12V$, $16.1V$, $19.5V$, and $30.5V$, which correspond to the Wigner crystal and Mott insulator states at n=1/3, 1/2, 2/3, and 1. These spectra are characteristic of electron doped graphene. They are comparable to the spectra at $V_{BG} = -2V$ and $0$ V for $V_{TG} = 0$ (shown in Fig. S1a). This indicates that the back gate can induce finite electron doping in the graphene when the $WS_2/WSe_2$ moiré heterostructure is in a correlated insulating state due to the electron incompressibility of the correlated insulators.

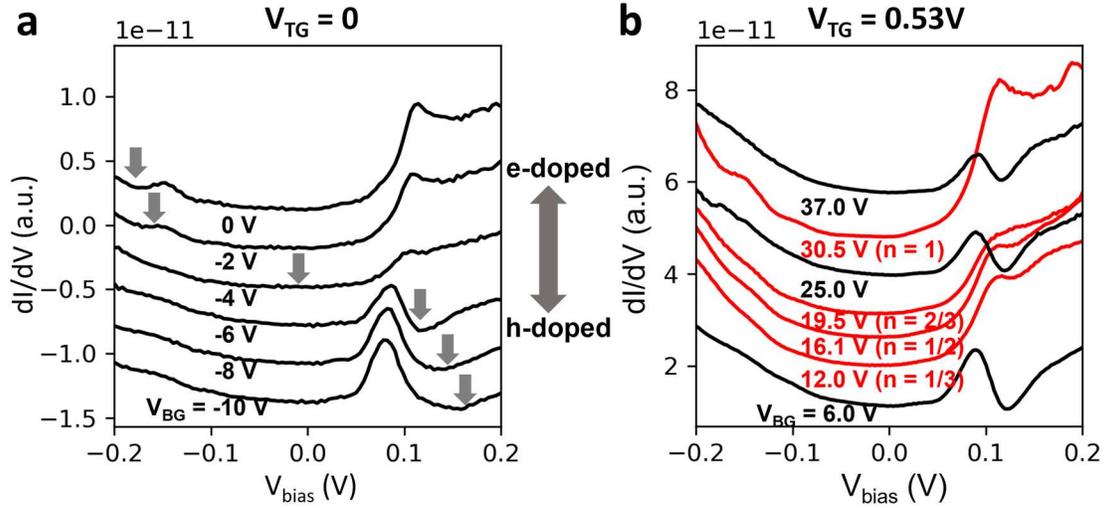

**Figure S1 Comparison of single dI/dV spectra obtained at $V_{TG} = 0$ and $V_{TG} = 0.53$ V. a**. $V_{TG} = 0$. **b**. $V_{TG} = 0.53$ V. In **a** we display dI/dV spectra obtained when the graphene doping is near the charge neutral point. A strong spectral change is observed when the graphene transitions from hole-doped to electron-doped. The Dirac point positions are denoted by vertical arrows. In **b** we display typical dI/dV spectra at n = 1/3, 1/2, 2/3, and 1 for correlated states (red) as well as for three other filling factors that lack correlated states (black). In each panel the dI/dV spectra are shifted vertically for clarity. The spectra indicate that the graphene sensing layer is more electron doped when the moiré heterostructure is in a correlated insulator state.

2. **Moiré site dependence of the dI/dV spectra**

Figure S2a shows the topography of the moiré superlattice. Figures S2b-d show the graphene dI/dV spectra along the AA-AB$_1$-AB$_2$-AA line in Fig. S2a for $V_{TG}$=0.7V and $V_{BG}$ = 19V, 26.5V, 35V, respectively. We find that the graphene dI/dV spectra are very similar at all positions. The doped $WS_2/WSe_2$ moiré heterostructure (and the Wigner crystal states) mostly shifts the average electron doping in the graphene monolayer. The spatial dependence is weak and only observable over narrow bias voltage ranges associated with the discharging of moiré electrons (as shown in Fig. 2 and Fig. 3 of the main text.)

We can understand this effect by noting that the doped moiré heterostructure provides only a relatively small modulation of the potential energy in graphene. Electrons in monolayer graphene are described by relativistic Dirac electrons which tend to be highly delocalized close to the charge neutral point. The screening length of Dirac electrons diverges at the charge neutral point and is longer than 8nm for doping level below $1 \times 10^{11} cm^{-2}$ (see ref. 2). As a result, electron density modulation (and associated dI/dV changes) at the moiré period scale is very small in charge neutral monolayer graphene.

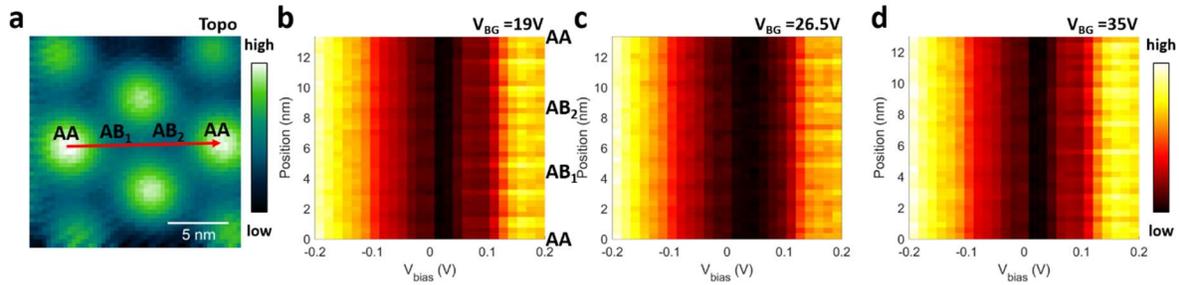

**Figure S2 Moiré site dependence of the dI/dV spectra**. **a**. A typical STM topographic image of the moiré superlattice seen through the graphene sensing layer. **b-d**. Position dependent dI/dV spectra measured along the red linecut shown in **a** with $V_{TG} = 0.7V$ and (**b**) $V_{BG} = 19V$, (**c**) 26.5V, and (**d**) 35V.

3. **Raw images and FFT filtering of the generalized Wigner crystal states**

The dI/dV images of the generalized Wigner crystal states shown in Fig. 2 of the main text are FFT filtered to suppress the periodic feature associated with the moiré superlattice. Figure S3 shows the unfiltered raw images of the generalized Wigner crystal states and the process of the FFT filtering.

Fig.S3a displays the raw dI/dV map of the n = 2/3 state, with the corresponding FFT image shown in Fig. S3b. Here the red circles label the positions of reciprocal unit vectors of the moiré superlattice. We filter out the signal within the red circles in the FFT filtering process. Fig. S3c shows the FFT filtered image of Fig. S3a (same as Fig. 2e in the main text) and Fig. S3d shows the corresponding filtered FFT image (same as Fig. 2f in the main text).

Similar filtering procedures are performed for the n = 1/3 and n = 1/2 Wigner crystal states. Figure S3e and S3f show the raw real space and FFT images, respectively, of the n=1/3 Wigner crystal state, and the corresponding FFT filtered images are shown in Fig. S3g and S3h. Figure S3i and S3j show the raw real space and FFT images, respectively, of the n=1/2 Wigner crystal state, and the corresponding FFT filtered images are shown in Fig. S3k and S3l.

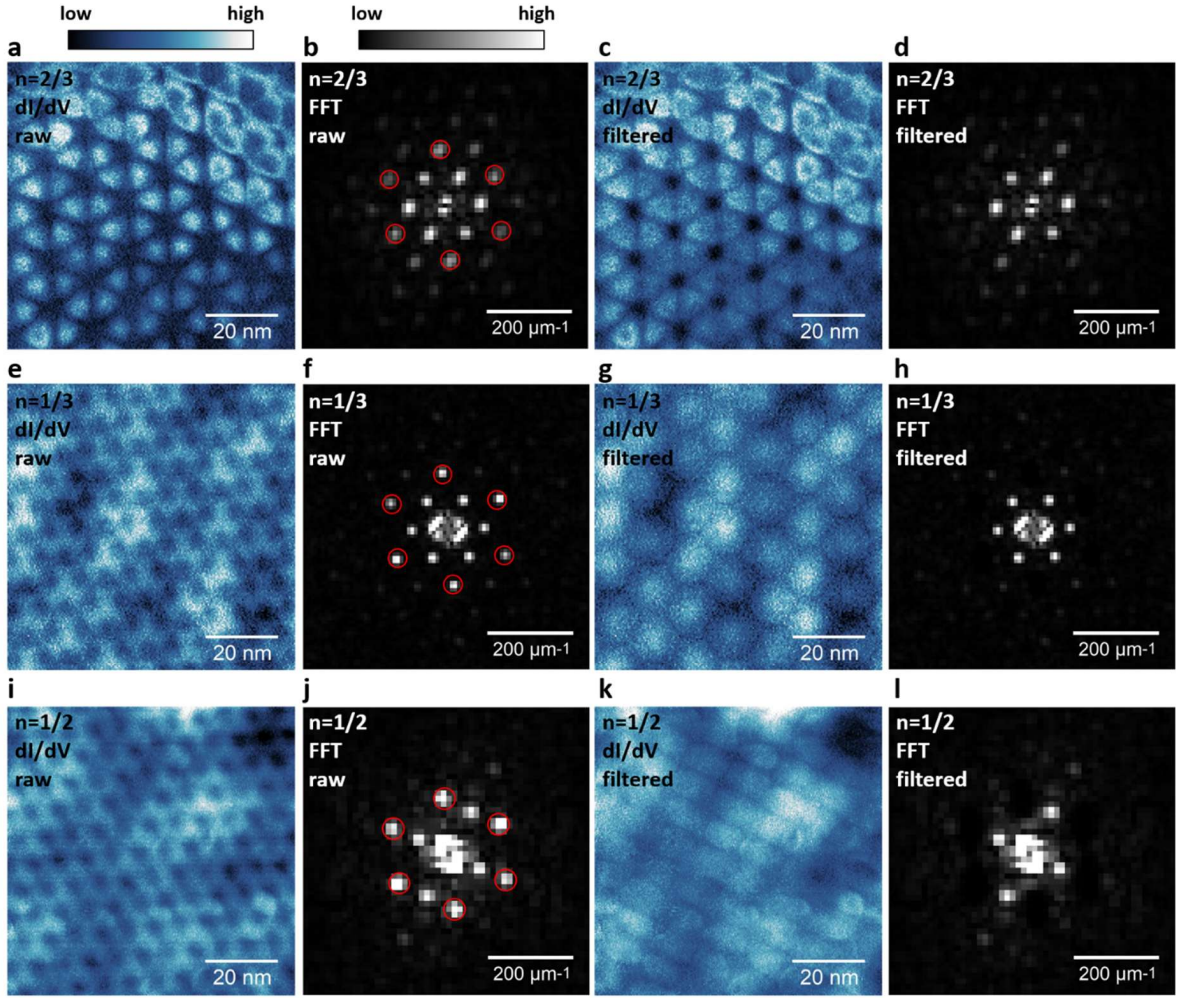

**Figure S3. Raw images and FFT filtering of the dI/dV maps for the generalized Wigner crystal states. a**. Raw dI/dV map of the n = 2/3 state. **b**. FFT image of (**a**). **c**. Real space dI/dV map after FFT filtering of (**a**). In the filtering process, we removed the Fourier components within the six red circles indicated in (**b**). This FFT filtering suppresses the periodic feature associated with the moiré superlattice. **d**. FFT image of (**c**). **e**. Raw dI/dV map of the n = 1/3 state. **f**. FFT image of (**e**). **g**. Real space dI/dV map after FFT filtering of (**e**). The Fourier components within the red circles shown in (**f**) have been filtered out. **h**. FFT image of (**g**). **i**. Raw dI/dV map of the n = 1/2 state. **j**. FFT image of (**i**). **k**. Real space dI/dV map after FFT filtering of (**i**). The Fourier components within the red circles shown in (**j**) have been filtered out. **l**. FFT image of (**k**).

**Reference:**

1    Zhang, Y. *et al.* Giant phonon-induced conductance in scanning tunnelling spectroscopy of gate-tunable graphene. *Nature Physics* **4**, 627-630 (2008).
2    Wong, D. *et al.* Spatially resolving density-dependent screening around a single charged atom in graphene. *Physical Review B* **95**, 205419 (2017).